# Using Ensemble Monte Carlo Methods to Evaluate Non-Equilibrium Green's Functions

**David K. Ferry**

School of Electrical, Computer, and Energy Engineering, Arizona State University, Tempe, AZ 25287-6206; ferry@asu.edu

## Abstract

The use of ensemble Monte Carlo (EMC) methods for the simulation of transport in semiconductor devices has become extensive over the past few decades. This method allows for simulation utilizing particles while addressing the full physics within the device, leaving the computational difficulties to the computer. More recently, the study of quantum mechanical effects within the devices, effects which also strongly affect the carrier transport itself, have become important. While particles have continued to be useful in quantum simulations using Wigner functions, interest in analytical solutions based upon the non-equilibrium Green's functions (NEGF) have become of greater interest in device simulation. While NEGF has been adopted by many commercial semiconductor, there remains considerable computational difficulty in this approach. Here, a particle approach to NEGF is discussed, and preliminary results presented illustrating the computational efficiency that remains with the use of particles. This approach adopts the natural basis functions for use in a high electric field and the preliminary results are obtained for quantum transport in Si at 300 K. This approach appears to offer significant advantages for the use of NEGF.

## 1. Introduction

In the classical world, transport in applied electric and magnetic fields is usually computed with the Boltzmann equation. In low fields, one has a nearly equilibrium situation in which the distribution function is a Maxwellian or a Fermi-Dirac at a temperature that likely increases above that of the lattice. In most devices, the distribution is well out of equilibrium, which means that it is unknown. Finding this distribution is typically the single most difficult problem. Classically, this entails solving the Boltzmann transport equation [1]. At least classically, an alternative approach is to use the computer to completely solve the transport problem with a stochastic methodology, and several methods have arisen to do this, two of which are an integral iteration technique [2,3] and the Monte Carlo method [4]. Today, the most widely used approach is the latter. The ensemble Monte Carlo (EMC) technique uses an ensemble of particles whose propagation is determined in parallel, so that ensemble averages can be computed as a function of time for the various observables of interest. It has been the subject of many reviews [5,6,7,8]. The methodology of the EMC approach is to spend time and effort on characterizing the correct physics, perhaps even to the use of atomistic full-band energy behavior [9], and the scattering properties, and let the computational burden be handled by a modern parallel processing computer. Hence, no attempt is made to arrive at an *ab initio* analytical form. These approaches have become extremely sophisticated, and many factors such as degeneracy [10], discrete impurities in real space [11], and many-body effects [12] can be incorporated.

From the beginning, there has been a rich history for the use of particles in quantum mechanics, as the suggestion of particles *and* waves dates to de Broglie [13], and the use of particles has been discussed over the years within the quantum world from Kennard [14] to Feynman [15]. The immediate problem with adapting the EMC approach for quantum transport lies with the use of the particle paths in phase space in the classical approach. Generally, the uncertainty principle restricts being able to simultaneously define both a position and momentum for a particle path. Nevertheless, this strict interpretation has been finessed by a variety of approaches, many of which clearly endorse some form of a real path for a particle that exists with the wave. Not the least of these approaches is due to Feynman himself. One more successful approach is the phase-space representation of the Wigner function [16], as this suggests a comparison between quantum dynamics and the corresponding classical motion [17,18]. However, the Wigner function can have a non-unitary evolution, and off-diagonal terms in the density matrix (from which the Wigner function is found) lead to non-classical propagation. Indeed, quantum coherence is reflected in oscillatory behavior of the Wigner function and non-positive definite regions in phase space, and this leads to complications in the use of EMC. Nevertheless, particle methods and the EMC have been adapted to treating quantum transport via the Wigner function [19,20,21].

Quite generally, the operation and performance of semiconductor devices has evolved to incorporate many quantum effects [22]. Any quantum mechanical simulation of such a device has to meet all the requirements of a self-contained and consistent mathematical theory, just as in the classical case. But it generally reflects a much deeper physical behavior. Certainly, a variety of quantum approaches to the simulation of transport have arisen, including the Wigner function mentioned above. Indeed, there have been many attempts to try to put quantum mechanics into the normal kinetic equations stemming from the Boltzmann approach, although most limit the method to approximating the spectral density that replaces the energy-conserving delta function for scattering [23,24,25,26]. Such an approach has also been taken with Bohmian trajectories and a quantum potential [27]. However, today one of the most popular approaches to quantum transport and quantum distributions is thought to be non-equilibrium Green's functions (NEGF). As in the classical case where the system is quite the same, the NEGF include new functions that must be found to describe both the non-equilibrium distribution function, but also how quantum correlations are built into the device [28,29]. These Green's functions also determine the transport of an "excitation" over a certain distance to where it is annihilated (thus these functions contain two times and two spatial variables). As a result, these functions bring considerable computational difficulty to any transport problem, with consequent long computation times. Nevertheless, they are being used extensively in modern device simulation and design.

It is known in quantum mechanics that any convenient coordinate system, or set of basis functions, may be used for any given problem, since the Schrödinger equation is a linear equation. In this paper, the use of a particular set of basis set is adopted that is extremely useful in a high electric field. These constitute the Airy transform, after which one can obtain an integral equation for the important "less-than" Green's function that is amenable to solution via EMC techniques. While the approach is currently restricted to uniform fields, and materials such as Si, it can be extended easily to the inhomogeneous world of semiconductor devices. In the following section, a brief introduction to NEGF, and the limitations of these functions, will be provided. Then, in Secs. 3 and 4, the Airy transform approach will be described and the various NEGF developed with this transform and the case of Si. The restrictions to Si will also be discussed. Section 5 will describe the EMC simulations and the results. Finally, in Sec. 6, these results will be discussed and prospects for extending the approach to inhomogeneous devices and to other materials will appear.

## 2. NEGF

In general, it is a well-known relationship in quantum mechanics that the wave function can be connected to initial conditions through the existence of a propagator as

$$\psi(\boldsymbol{x},t) = \int d\boldsymbol{x}' K(\boldsymbol{x},\boldsymbol{x}';t,0)\psi(\boldsymbol{x}',0) \, , \qquad (1)$$

where one can write the propagator as

$$K(\boldsymbol{x},\boldsymbol{x}';t,0) = \sum_i \varphi_i^\dagger(\boldsymbol{x}')\varphi_i(\boldsymbol{x})\, e^{-E_i t/\hbar} \, , \qquad (2)$$

and the $\varphi_i(\boldsymbol{x})$ are basis functions in a spatial expansion of the total wave function, where $E_i$ are the eigenvalues for the basis states (although in the following use of Airy functions, only the total energy will be important). It was Schwinger who formally connected this propagator to Green's functions, which were already well-known in the mathematics of differential equations [30]. Initially, these were independent of time, as he was concerned with the equilibrium state, and he assumed time reversibility, hence sticking to an equilibrium system. This equilibrium was maintained, even in the presence of perturbations, by the equality of $t \to \pm\infty$. These perturbations were handled formally by the existence of a unitary operator that involved the perturbing Hamiltonian exponentially [31], although this operator was developed much earlier [32]. The important Green's functions in this equilibrium theory were the advanced and retarded functions given as

$$\begin{aligned} G_a(r,r';t,t') &= i\Theta(t-t') \\ &\quad \times \langle\{\psi^\dagger(r',t'),\psi(r,t)\}\rangle \\ G_r(r,r';t,t') &= -i\Theta(t-t') \\ &\quad \times \langle\{\psi(r,t),\psi^\dagger(r',t')\}\rangle \end{aligned} \, . \qquad (3)$$

Here, the curly brackets denote the normal anti-commutator relationships for fermions, while the angular brackets define normalization conditions, obtained with the equilibrium distribution.

It is a well-recognized fact that the application of an electric field leads to current and a phase-transition that breaks the time-reversal symmetry of these functions [33]. This then leads to the non-equilibrium state. Normalization cannot be done with the equilibrium distribution and one needs two additional Green's functions, the correlation functions, in order to gain the non-equilibrium distribution. These correlation functions are given by

$$G^{<}(\boldsymbol{x},\boldsymbol{x}';t,t') = i\langle\hat{\psi}^{\dagger}(\boldsymbol{x}',t')\hat{\psi}(\boldsymbol{x},t)\rangle \\ G^{>}(\boldsymbol{x},\boldsymbol{x}';t,t') = -i\langle\hat{\psi}(\boldsymbol{x},t)\hat{\psi}^{\dagger}(\boldsymbol{x}',t')\rangle \, , \quad (4)$$

and are known as the "less than" and "greater than" Green's functions, respectively. These are direct wave function products and do not have the anti-commutators that appear in (3). The important one for our purposes is the less-than function, which has an important relation to the distribution function via

$$G^{<}(k,\omega) = f(\omega)A(k,\omega) \, , \quad (5)$$

which is known as the Kadanoff-Baym ansatz [29]. Note that here, the Green's function now depends upon a single momentum and a single frequency. This implies explicitly that, in the homogeneous material to be considered, only the differences $x - \boldsymbol{x}'$ and $t - t'$ are important (and not the averages $(\boldsymbol{x} + \boldsymbol{x}')/2$ and $(t + t')/2$). In (5), $f(\omega)$ is the non-equilibrium distribution, which evolves from the equilibrium Fermi-Dirac function

$$f_{FD}(\omega) = \left[1 + e^{(\hbar\omega - E_F)/k_BT}\right]^{-1} \, , \quad (5a)$$

where $E_F$ is the Fermi energy.

There are some views that (5) does not satisfy causality, and another form is required [34]. However, precisely this form will be found with the Airy transforms and will allow the determination of an equation for the distribution function $f(\omega)$. The spectral density $A(k,\omega)$ retains exactly the same form as in equilibrium and is connected to the retarded and advanced functions through

$$A(k,\omega) = -Im\{G_r(k,\omega) - G_a(k,\omega)\} \, . \quad (6)$$

Each of these functions requires a pair of equations of motion. With the added complications of having four functions, these equations will be far more complicated than the simple Boltzmann equation. Hence, solving for these functions, and obtaining the distribution function in the quantum case, is a difficult problem, both analytically and computationally [22]. Indeed, it would be much easier if an approach based, for example, on EMC could be found. But, the computational problems are not the only problems with NEGF, for the equations needed are not limited to just these functions. In the quantum transport situation, one must deal with the existence of long-range correlations among the carriers, correlations that are not always alleviated with scattering. An example is given by impurity scattering, which is very long range due to the Coulomb interaction potential. Even classically, it has been observed that an electron can be interacting with several impurities at the same time [35]. In addition, with the wave accompanying the particle, a single Coulomb scattering center can create interference, much like a two-slit experiment, a result that has been found with both Green's functions [36] and Wigner functions [37]. These correlations affect the quantum transport and can lead to observable fluctuations in devices [38]. Such long range correlations are not limited to impurity scattering. They are expected for any long range potential; polar optical scattering is also Coulombic in nature and can be expected to lead to the same effects. Consequently, such correlations lead to difficulties when these scatterers are introduced perturbatively, as they require minimally evaluation of the Bethe-Salpeter equation for determination of mobility and transport [22,39]. This raises the difficulty level by a significant amount [40].

However, it has generally been believed that phonon scattering is "phase breaking" [41]. In such a situation, long-range correlations that exist in the phase are broken, and there may be no need to proceed to the inclusion of the Bethe-Salpeter equation. This is certainly likely the case for non-polar scattering, which is assumed here. But, it may also be true for polar-mode scattering, and more work is needed to ascertain whether this is true or not.

### 3. Airy Transforms

One of the first requirements in quantum transport is the choice of wave functions and the corresponding basis set. In the presence of an electric field that is taken along the z-direction for convenience, the one-dimensional eigenfunctions along this direction are determined from (time variation is assumed to enter in the normal manner, and only the *z*-variation is considered at the moment)

$$\left(-\frac{\hbar^2}{2m^*}\frac{\partial^2}{\partial z^2} + eFz\right)\varphi(z) = E\varphi(z) \, , \quad (7)$$

where the symbol $F$ is used to denote the field in order to differentiate it from the energy $E$. The wave functions in the two orthogonal directions $(x,y)$ are taken to be normal plane waves that exist in the transverse band structure. For this latter, the normal $\boldsymbol{k}$ and $\boldsymbol{r}$ will denote the wave vector and position in the transverse plane of $(x,y)$-coordinates. As may be found in many quantum mechanics textbooks, the solutions to (7) are the Airy functions [42]

$$\varphi(\boldsymbol{r},z) = \frac{1}{2\pi L}e^{i\boldsymbol{k}\cdot\boldsymbol{r}}Ai\left(-\frac{z-z_0}{L}\right) , \quad (8)$$

where $z_0$ is an appropriate zero of the Airy function $Ai(\cdot)$. The factor $L$ is a normalizing length and is given by

$$L = \left(\frac{\hbar^2}{2m^* eF}\right)^{1/3}, \quad (9)$$

where $m^*$ is the effective mass. The basis set of Airy functions has been used to find the analytical eigen-values for the triangular potential wells in a MOSFET for many years [42]. In the MOSFET case, this provides the set of discrete bound state eigen-values and eigenfunctions determined by the discrete zeroes of the Airy function [41]. *This is not the case of interest here*.

In the present situation, a continuous *Airy transform* is to be used. The difference is exactly the same as the difference between a Fourier series and a Fourier transform. The former is a discrete set while the latter is a continuous function. In the present case, $s$ well be the continuous transform variable and the discrete set of zeroes will not appear explicitly. This transform is defined just as a Fourier transform would be defined. The Airy transform of the function $f(\boldsymbol{r}, z)$ is given as

$$F(\boldsymbol{k}, s) = \int d^2\boldsymbol{r} \int \frac{dz}{2\pi L} e^{i\boldsymbol{k}\cdot\boldsymbol{r}} Ai\left(\frac{z-s}{L}\right) f(\boldsymbol{r}, z) \,. \quad (10)$$

For the Green's function, there are two spatial variables, and hence one needs a double Airy transform, although it is fair to assume homogeneity in the transverse dimensions. This transform may then be expressed as

$$F(\boldsymbol{k}, s) = \int d^2\boldsymbol{r} \int \frac{dz}{2\pi L} \int \frac{dz'}{2\pi L} e^{i\boldsymbol{k}\cdot\boldsymbol{r}} \times Ai\left(\frac{z-s}{L}\right) Ai\left(\frac{z'-s'}{L}\right) f(\boldsymbol{r}, z, z') \,. \quad (11)$$

Here, the Green's function response is only being developed for the one dimension along the electric field. One could, of course, also extend the equations to two variables in the transverse directions, but here only the one will be used to ease the complexity of the resulting equations. A function that is diagonal in both $\boldsymbol{k}$ and $s$ is translationally invariant in the transverse plane, but not necessarily in the longitudinal plane along the electric field. In addition, a function that depends only upon the difference $z - z'$ will yield a transform that depends only upon $s = s'$.

## 4. Airy Form of NEGF

The development of Green's function transport using the Airy functions is relatively old [43,44,45,46], and the present treatment is focused upon the development of the EMC method of its evaluation. Consequently, the details of various derivations will be omitted here, but can also be found in [40].

### *4.1 The Retarded and Advanced Functions*

In the transverse dimensions, the Green's function is translationally invariant and so is a function only of $\boldsymbol{x} - \boldsymbol{x}'$. Solving the differential equation for the retarded and advanced functions leads to an integration over the intermediate variables which appears as a convolution integral. After the transforms are taken, the convolution becomes a simple product, so that the retarded Green's function satisfies the inhomogeneous integral equation [44]

$$\begin{aligned} G_r(\boldsymbol{k}, u, u') &= G_0^F(\boldsymbol{k}, u, u') \\ &+ \int du_1 \int du_2 G_0^F(\boldsymbol{k}, u, u_1) \\ &\times \Sigma_r(\boldsymbol{k}, u_1, u_2) G_r(\boldsymbol{k}, u_2, u') \end{aligned} \,. \quad (12)$$

Here, the short-hand notation $u = (z, t)$ has been introduced. In fact, in the case of significant correlation among the carriers, the argument of the integrals could properly be considered a three-particle Green's function, which would be incredibly difficult to evaluate [15,47,48]. In most cases, it is assumed that it can be rewritten as the product of three single-particle functions as shown. These can also lead to disconnected diagrams which cannot be discarded in the NEGF cases [32,49], as these generate important phase factors that affect any correlation and/or interference terms [50]. These higher-order effects must be considered and will be discussed in the Si model. The unperturbed Green's function, the first term on the right-hand side of (12), includes the role of the electric field and may be taken to be [51]

$$\begin{aligned} G_0^F(\boldsymbol{k}, s, s', t, t') &= G_{r,0}^F(\boldsymbol{k}, s, s', t, t') \\ &= -\frac{i}{\hbar}\theta_0(t - t') \\ &\times e^{-iE(\boldsymbol{k},s)(t-t')/\hbar} \delta(s - s') \end{aligned} \,. \quad (13)$$

The function $\theta_0(\xi)$ is the normal Heavyside step function which is 1 for $\xi \geq 0$, and 0 otherwise. Here, the energy is composed of the transverse energy plus that induced in the $z$-direction by the electric field, and this energy may be written as

$$E_{\boldsymbol{k},s} = E(\boldsymbol{k}, s) = \frac{\hbar^2 k^2}{2m^*} + eFs \,. \quad (14)$$

Both forms for the energy will be used below. The temporal Fourier transform of (13) is easily taken to be

$$G_0^F(\boldsymbol{k}, s, s', \omega) = \frac{\delta(s - s')}{\hbar\omega - E(\boldsymbol{k}, s) + i\eta} \,. \quad (15)$$

where the limit $\eta \to 0$ is usually taken, just as in the equilibrium Green's function case. Here, the convergence factor is confusing, since it implies the field was turned on in the infinite past. But, this is not the case. As may be seen in (13), the field is turned on at the time $t'$, so this convergence factor is useful, but not the correct interpretation. Care has to be taken over this point, but it will not affect the final results.

The full transformation with the Airy functions can now be taken to yield the equation for the retarded Green's function to be [44]

$$G_r(\boldsymbol{k}, s, s', \omega - E_{\boldsymbol{k},s}) = G_0^F(\boldsymbol{k}, s, \omega)\delta(s - s') \\ \times \int \frac{d^3\boldsymbol{q}}{8\pi^2} \sum_{\nu=\pm1} |M_q|^2 \left(N_q + \frac{1+\nu}{2}\right) \delta(\kappa) \\ \times G_r(\boldsymbol{w}, \boldsymbol{w}', \kappa') \tag{16}$$

with the definitions and substitutions

$$\kappa = \omega - \nu\omega_0 - E_{\boldsymbol{k},s} + E_{\boldsymbol{k}',s} \\ \kappa' = \omega - \nu\omega_0 - E_{\boldsymbol{k},s'} \\ \boldsymbol{w} = \boldsymbol{k} + k_z \boldsymbol{a}_z + \nu \boldsymbol{q} \\ \boldsymbol{w}' = \boldsymbol{k}' + k_z' \boldsymbol{a}_z + \nu \boldsymbol{q} \\ N_q = \left[e^{\hbar\omega_0/k_B T} - 1\right]^{-1} \tag{17}$$

where the last line is the Bose-Einstein distribution.

The retarded self-energy is the product of the retarded Green's function and the retarded phonon Green's function. In most semiconductors, the scattering is weak, and the self-energy can be calculated to lowest order in time-dependent perturbation theory. This means that the Green's function propagator is approximated by the free propagator (in the presence of the field) (15). While this may suggest that physical effects of the scattering are being minimized, this isn't the real case. The retarded Green's function (16) is still being solved self-consistently, and the self-energy corrections will certainly be built into the result from Dyson's equation. Here, a non-degenerate electron gas will be considered as discussed below. This is equivalent to neglecting any screening of the optical phonons. Moreover, the phonons are taken to be in equilibrium, so that the normal Bose-Einstein distribution, at the lattice temperature, can be used to describe the statistics of the phonons. The retarded phonon propagator is then

$$D_{0,r}(\boldsymbol{q}) = -i\pi \sum_{\nu=\pm1} |M_q|^2 \\ \times \left(N_q + \frac{1+\nu}{2}\right) \delta(\kappa) \tag{18}$$

The summation runs over emission and absorption of the optical phonon, with the phonon energy considered to be constant, $M_q$ is the electron-phonon matrix element, and $N_q$ is the Bose-Einstein distribution. This leads to the self-energy term as

$$\Sigma_r(\boldsymbol{k}, \boldsymbol{k}', z, z', \omega) = \int \frac{d^3\boldsymbol{q}}{8\pi^2} \sum_{\nu=\pm1} |M_q|^2 \\ \times \left(N_q + \frac{1+\nu}{2}\right) \delta(\kappa) G_r(\boldsymbol{w}, \boldsymbol{w}', \kappa') \tag{19}$$

While this still remains relatively simple, it is now necessary to take the Airy transform of this function, and this provides adequate complication. The Airy transform (11) is applied to both longitudinal variables, and may be expressed as

$$\Sigma_r(\boldsymbol{k}, s, s', \omega) = \int \frac{d^3\boldsymbol{q}}{8\pi^2 L^4} \sum_{\nu=\pm1} |M_q|^2 \left(N_q + \frac{1+\nu}{2}\right) \\ \times \int ds_1 \int ds_2\, G_r(\boldsymbol{k} + \nu\boldsymbol{q}, s_1, s_2, \kappa) \\ \times \int ds_1 \int ds_2 \int dz \int dz' Ai(z - s) \\ \times Ai(z - s_1) Ai(z' - s') Ai(z' - s_2) \tag{20}$$

Here, the Airy functions have reduced arguments with the values

$$Ai(y) = Ai\left(\frac{eFy - \hbar\omega}{\Theta}\right) \\ \Theta = \left[\frac{3(e\hbar F)^2}{2m^*}\right]^{1/3} = eFL \tag{21}$$

Because $G_0^F$ is diagonal in *s*, it may be expected that the resultant self-energy is also going to be diagonal in *s* once the transform has been taken. Then, the integrals can be performed using the known properties of the Airy functions and their integrals [52,53,54,55,56]. The resulting self-energy is then found to be [44]

$$\Sigma_r(\boldsymbol{k}, s, \omega) = \frac{(m^*)^{3/2} |M_q|^2 \sqrt{\Theta}}{\sqrt{2}\hbar^2} \\ \times \sum_{\nu=\pm1} \left(N_q + \frac{1+\nu}{2}\right) F(s, \kappa) \tag{22}$$

with

$$Im\{F(s, \kappa)\} = Ai'(-y)Bi'(-y). \\ -yAi(-y)Bi(-y) \\ Re\{F(s, \kappa)\} = Ai'^2(-y) - yAi^2(-y) \tag{23}$$

where *Bi* is the Airy function of the second kind, and the primes denote derivatives with respect to the argument. These functions will be plotted below, once the material system has been described.

### *4.2 The Correlation Functions*

The less-than Green's function will originate from the above retarded functions, and will satisfy its own set of differential equations. From this less-than function, a distribution function will evolve. This is the goal, to find the quantum non-equilibrium distribution function. This function will satisfy an integral equation that will be amenable to EMC. The differential equations to be solved are Fourier and Airy transformed to become

$$[\hbar\omega - E_{\boldsymbol{k},s}] G^<(\boldsymbol{k}, s, s', \omega) = \\ = [\Sigma_r(\boldsymbol{k}, s, \omega) G^<(\boldsymbol{k}, s', \omega) \\ + \Sigma^<(\boldsymbol{k}, s, \omega) G_a(\boldsymbol{k}, s', \omega)] \tag{24}$$

and

$$[\hbar\omega - E_{\boldsymbol{k},s'}] G^<(\boldsymbol{k}, s, s', \omega) =. \\ = [G_r(\boldsymbol{k}, s, \omega) \Sigma^<(\boldsymbol{k}, s', \omega) \\ + G^<(\boldsymbol{k}, s, \omega) \Sigma_a(\boldsymbol{k}, s', \omega)] \tag{25}$$

Normally, at this point, the sum and difference of these two equations are taken. Here, however, this leads to the same equation [45], which may be directly solved for the less-than function. Taking the sum of the two equations, and performing a little algebraic reorganization, one finds that the less-than function reduces simply to [45]

$$G^<(\boldsymbol{k},s,s',\omega) = G_r(\boldsymbol{k},s,\omega) \times G_a(\boldsymbol{k},s',\omega)\Sigma^<(\boldsymbol{k},s,s',\omega) . \quad (26)$$

It is easily shown that the difference of the two equations leads to the same result. One can easily also show that $G_r - G_a = 2Im\{\Sigma_r(\boldsymbol{k},s,\omega)\}G_rG_a$. These constraints allow us to rewrite the less-than function in the generalized Kadanoff-Baym manner as

$$G^<(\boldsymbol{k},s,\omega) = A(\boldsymbol{k},s,\omega)f(s,\omega) , \quad f(s,\omega) = \frac{\Sigma^<(s,,\omega)}{2Im\{\Sigma_r(s,\omega)\}} . \quad (27)$$

Here, $f(s,\omega)$ is the full non-equilibrium distribution function. In this form, a quantum distribution function has been defined and is clearly not one of equilibrium. Normally, in the Kadanoff-Baym form (5), the distribution function is only a function of $\omega$, but the parameter $s$ has been retained here to indicate the dependence of the various paths for the particles on this quantity. While the form of this distribution function appears to be simple enough, care must be taken because the details of the less-than self-energy will certainly make this equation difficult.

Just as the retarded self-energy was developed, the less-than self-energy can be easily developed for the case of the non-polar intervalley phonons in Si. The self-energy can be written in the Airy transformed variables as

$$\Sigma^<(\boldsymbol{k},s,\omega) = \frac{|M_q|^2}{3^{1/6}L^2}\sum_{\nu=\pm1}\left(N_q + \frac{\nu+1}{2}\right) \times \int d^2\boldsymbol{q}_t \times \int ds' Ai^2\left(\frac{s-s'}{3^{1/3}L}\right) \times G^<(\boldsymbol{k}-\nu\boldsymbol{q}_t,s',\omega-\nu\omega_0) . \quad (28)$$

As this latter equation depends upon the less-than Green's function, it is clear that it will lead to an integral equation for the distribution function, that appears in (27). The integration over $\boldsymbol{q}_t$ removes all dependence on the transverse momentum as it leads to a conservation of this momentum (all uncertainty now exists in the longitudinal change in momentum and this is in the spatial Airy variables. As a result, the distribution function can be written as the integral equation [45**Error! Bookmark not defined.**]

$$f(s,\omega) = \frac{1}{Im\{\Sigma_r(s,\omega)\}}\sum_{\nu=\pm1}\left(N_q + \frac{\nu+1}{2}\right) \times \int ds' K(s,s',\omega-\nu\omega_0)(s',\omega-\nu\omega_0) \quad (29)$$

with

$$K(s,s',\omega) = \frac{\sqrt{3}|M_q|^2 m^*}{\hbar L^2} Ai^2\left(\frac{s-s'}{3^{1/3}L}\right)\left[\frac{\pi}{2} + atanh\left(\frac{\hbar\omega - E_{\boldsymbol{k},s'} - Re\{\Sigma_r(,s',\omega)\}}{Im\{\Sigma_r(,s',\omega)\}}\right)\right] . \quad (30)$$

While it may not seem evident, this integral equation for the distribution function may be readily solved with EMC techniques. In the classical approach, the ballistic drift of the carriers is for a period of time that is determined from the probability that a carrier has not been scattering over that period of time. This is represented by the probability being a negative exponential of the time relative to the inverse of the scattering rate. The same principle applies. However, instead of the time, it is the ballistic distance $s - s'$, relative to the distance $L$. From this distance, one can determine the ballistic time and other dynamic variables, as discussed below. Thus, one has the same iterated procedure as in classical EMC: drift for a period of time/space, then scattering according to the various processes in the less-than self-energy. The leading term involving the imaginary part of the retarded self-energy serves merely to renormalize the distribution.

### *4.3 The Si Model*

To illustrate this new EMC approach to solving for NEGF, the case of Si will be considered. While the situation for electrons in the conduction band is complicated by the multi-valley nature of this band, the scattering processes themselves are very local and do not require some of the more complicated higher-order corrections to the simpler Fermi golden rule, although the latter is of course different for the quantum case. Although Si is the most common semiconductor known, and is the basis for the entire semiconductor industry, some of its intrinsic properties are not as well known. Here, a few of these are given, and a more extensive discussion is presented in the Appendix below. The conduction band of Si has six equivalent ellipsoids located along the Δ lines [these are the set of (100) axes] about 85% of the way to the zone edge at X. Since scattering between these equivalent valleys is possible, the electric field will be taken along the (111) direction so that the same angle is made with each of the six valleys. Scattering within each ellipsoid is limited to acoustic phonons, as intra-valley optical processes are forbidden. Acoustic mode scattering, by way of the deformation potential, is characterized by two constants $\Xi_u$ and $\Xi_d$, which are thought to have values of 9 eV and -6 eV, respectively [57]. The effective deformation potential is then the sum of these, or about 3 eV. Non-polar "optical" phonon scattering occurs for scattering between the equivalent ellipsoids. There are two possible types of phonons that are involved in this process (these phonons are discussed further in the appendix). One is the *g*-phonon that couples the two valleys along opposite ends of the same (100) axis. This is an *umklapp* process so that, after reduction by a reciprocal lattice vector, the scattering has a net phonon wave vector of $0.3\pi/a$ (see figure A3). The symmetry allows only the LO mode to contribute to this scattering. At the same time, *f*-phonons couple the (100) valley to the (010) and (001) valleys, and so on. The latter wave vector lies in the (110) direction has a magnitude of $2^{1/2}(0.85)\pi/a = 1.2\pi/a$, which lies in the square face of the Brillouin zone along the extension of the (110) line into the second Brillouin zone [58]. The phonons here are near the X-point phonons in value but have a different symmetry.

The energies of the LO *g*-phonon and the LA and TO *f*-phonons have nearly the same value, while the low-energy inter-valley phonons are forbidden. Long [59], however, determined from an analysis of the experimental mobility versus temperature that a weak low-energy inter-valley phonon is required to fit the experimental data. He treated the high-energy phonons by a single equivalent inter-valley phonon of 64.3 meV, but introduced a low-energy inter-valley phonon with an energy of 16.4 meV. The presence of the low-energy phonons is also confirmed by studies of magneto-phonon resonance (where the phonon frequency is equal to a multiple of the cyclotron frequency) in Si inversion layers, which indicates that scattering by the low energy phonons is a weak contributor to the transport [60]. The low-energy phonon is certainly forbidden although Long treats it with a very weak coupling constant. It is more likely that this forbidden low-energy inter-valley phonon must be treated by a first-order interaction [61,62]. This fits the data with a coupling constant of $\Xi_0 = 5.6$ eV, while the allowed, higher-energy transition is treated with a coupling constant $D = 9 \times 10^8$ eV/cm. This two-phonon model is adopted for the present treatment of Si.

Non-parabolicity of the conduction bands away from the $\Delta$ minima arises more from repulsion from the upper (second) conduction along these directions than from a valence band interaction. Non-parabolicity is included in the simulations by the normal hyperbolic bands with an effective "gap" of 2.1 eV [58]. This non-parabolicity will be important in computing velocities from the energies and momentum determined during the Monte Carlo process described in the next section.

The various scattering rates that go into the retarded self-energy, under the approximations described above, and for a field of 10 kV/cm, are shown in figure 1. An unusual part is that the "scattering" rates begin for negative energy in some cases involving the phonon absorption processes.

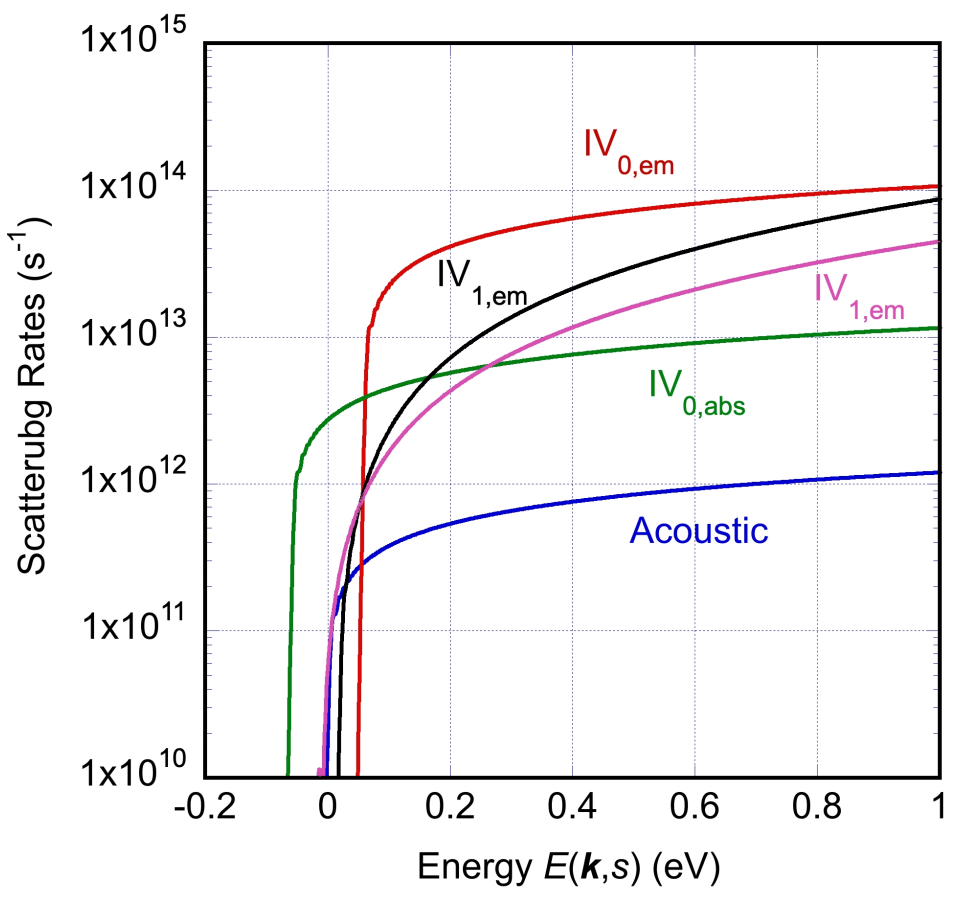

**Figure 1**. The scattering rates that enter the retarded self-energy are shown for Si at room temperature. These include the zero- and first-order intervalley optical phonons and the acoustic phonons. The rates are for a field of 10 kV/cm.

Normally, in classical situations, they occur only for positive energies due to the vanishing of the density of final states which is zero for negative energy. Here, however, the form of the retarded self-energy in (19) involves a Green's function, rather than the density of states, that has values at negative energy due to the particular forms of (17).

The real and imaginary parts of the retarded self-energy are shown in figure 2 for three values of the electric field (10, 40, 70 kV/cm). The oscillations in the real part lead to the quasi-quantization of the density of states into two-dimensional sub-bands in the high electric field. The field tends to create a triangular potential well, which is expected from the quantum mechanics. When the real part is positive, the energy is lowered, while when it is negative, the energy is raised. This tends to form a set of sub-band energy levels [63]. This tendency toward formation of sub-bands is also evident in the imaginary part of the self-energy. It is clear, however, that the general shape of the imaginary part is only weakly dependent on the value of the electric field. The zero of energy in these plots is taken as the conduction band minimum for the one-electron bands. Broadening arises from both the scattering and the field, and this leads to values for negative energy.

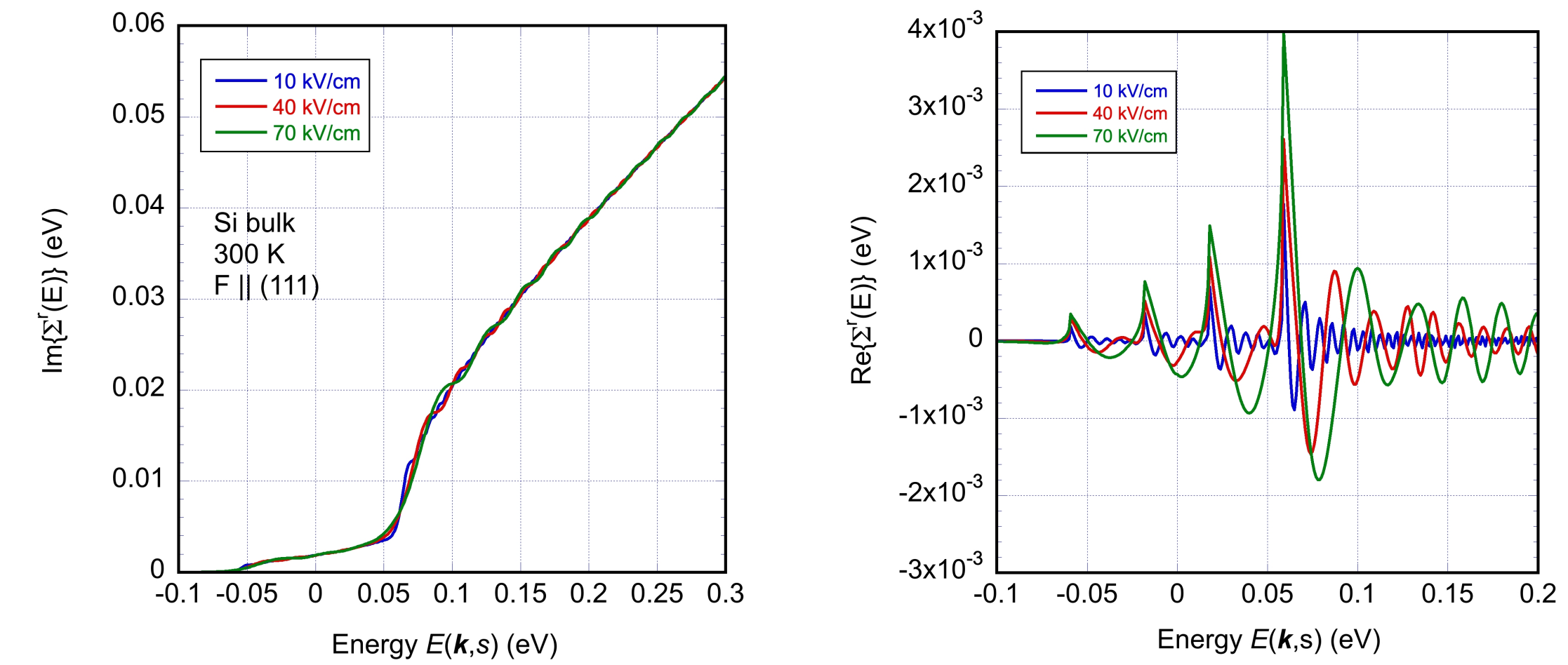

(a) (b)

**Figure 2.** (a) The imaginary part of the retarded self-energy as a function of the energy itself. (b) The real part of the retarded self-energy as a function of the energy itself. These are determined using the scattering mechanisms described in the text for Si at 300 K and for three values of the electric field (10, 40, and 70 kV/cm).

The retarded self-energy leads to the spectral density which describes how the various energy levels are broadened by the scattering and quantization processes. This spectral density is shown in figure 3 for the three field values used in illustrating the self-energy. Here, it appears that, as with the imaginary part of the self-energy, the field has only a small effect on the spectral density. Thus, it may be concluded that the retarded (and advanced) functions are not particularly dependent upon the actual value of the electric field applied to the device, other than the changes due to the tendency to form two-dimensional sub-bands. These functions are more related to the intrinsic properties of the material used in the simulations, and not so much upon the actual value of the external field.

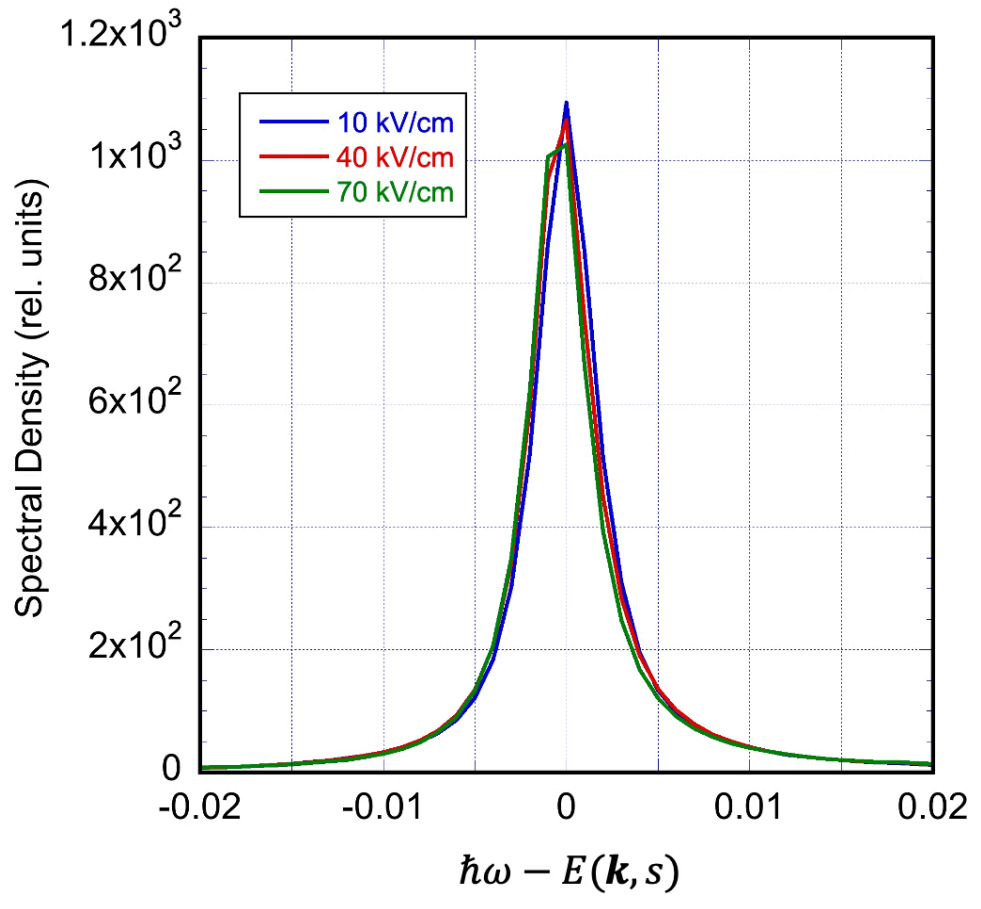

**Figure 3**. The spectral density obtained from the retarded (and advanced) self-energies for Si at 300 K.

## 5. The EMC Results

Following from the above discussion, and the results of (29) and (30), the sequence of the Monte Carlo steps follow a similar path to that for the Boltzmann equation. Here, however, the exponential in time is replaced by the Airy function in space. That is, drift is defined by an effective path length first, and then scattering occurs. The difference between this approach and the Monte Carlo for the Boltzmann equation is that drift length replaces drift time, although these are intimately related, as will be described below. It may be noted that the imaginary part of the retarded self-energy sits outside the integral and is a final adjustment on the distribution function. To begin, the spatial motion of the particle is determined with a random number $r$ according to

$$r = \int_{-u_0}^{u} Ai^2(u')du' \Big/ \int_{-u_0}^{\infty} Ai^2(u')du' \ , \qquad (31)$$

so that a value for the drift length is determined from $u$ as

$$\Delta s = 3^{1/3} L u \ . \qquad (32)$$

There is a question about the use of the Airy function because it has values over a semi-infinite range of values. However, one may note that the scattering function in (30) seems to be over a single effective sub-band (the values of the bracketed term), so that the quantity $-u_0$ is taken to be the first zero of the Airy function, so that there is a single maximum in the quantity integrated in (31). This gives a range of values for $\Delta s$ for the ensemble of particles. This may be averaged over the ensemble and over the number of iterations to determine an "average" value of $\Delta s$, and this is plotted as a function of the electric field in figure 4. The error is extremely small, so that error bars do not show up; the average is computed over the $10^5$ particles and for 200 iterations of the EMC algorithm. While the velocity appears to converge rapidly, as discussed below, the number of iterations is used to assure that the distribution function is converged.

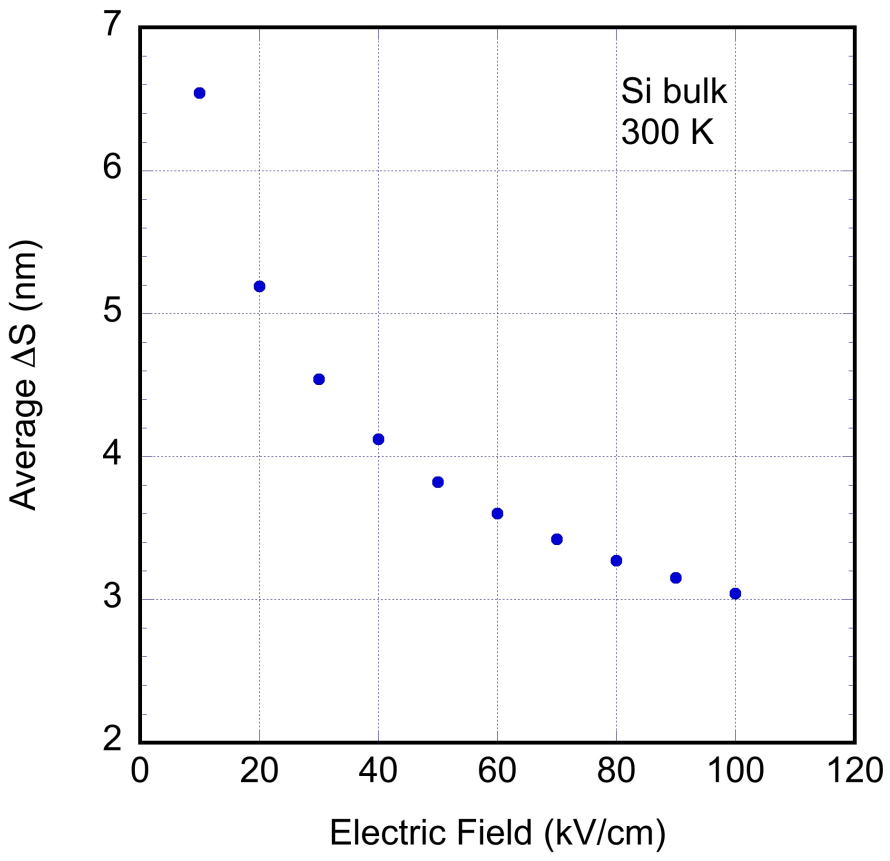

**Figure 4.** The average drift length Δs as a function of the electric field. The average is computed over $10^5$ particles and 200 iterations of the Monte Carlo algorithm.

In the EMC procedure, each particle has its own value for $\Delta s$. This leads to an energy gain in the field of $eF\Delta s$, which is directed along the electric field (taken to be the $z$ axis for convenience). As a result, the momentum wave vector is changed by an amount $\Delta k = (eF/\hbar)\Delta t$ (but see a comment below). It is therefore necessary to find this value for $\Delta t$. The drift time and the drift distance are trivially related in classical mechanics where the bands are parabolic. With the non-parabolic bands that exist in Si, this is no longer the case, and some work is necessary to connect these two quantities. The problem arises from the energy dependence of the effective mass in the non-parabolic bands, which breaks the simple connection between momentum wave number and velocity. In the hyperbolic bands used, the effective mass increases with energy according to [58]

$$m^* = m_c(1 + E/E_G) \, , \qquad (33)$$

where $m_c$ is the mass at the conduction band minimum and $E_G$ is the effective energy gap mentioned in the previous section. This is determined for the one-electron bands, as the broadening is added with the quantization in the Green's functions. The change in velocity can be written as

$$\Delta v(t) = \frac{eFt}{m_c}\left(1 + \frac{E(0)}{E_G} + \frac{eFvt}{E_G}\right)^{-1} , \qquad (34)$$

which leads to a difficult integral equation. Here, $E(0)$ is the energy at the start of the drift. We may take the distance as $z = vt$. The latter suggests that the velocity be written as $dz/dt$, so that (34) may be rewritten as

$$\frac{dz}{dt}\left(1 + \frac{E(0)}{E_G} + \frac{eFz}{E_G}\right) = \frac{eF}{m_c} t \, , \qquad (35)$$

with $z$ running from 0 to $\Delta s$ and $t$ running from 0 to $\Delta t$. This integral is now trivially evaluated to lead to the value for the increment in time as

$$\begin{aligned} \Delta t &= \sqrt{\frac{2m_c}{eF}|\Delta s|\left(1 + \frac{E(0)}{E_G} + \frac{eF\Delta s}{2E_G}\right)} \\ &= \sqrt{\frac{2m_c}{eF}|\Delta s|\left(1 + \frac{E}{E_G} - \frac{eF\Delta s}{2E_G}\right)} \end{aligned} \, , \qquad (34)$$

where $E$ is now the energy at the end of the drift. Once this time increment is determined, the momentum wave vector can be updated, and the velocity determined at the end of the drift period. It also allows one to determine the average drift time for the ensemble of carriers, in order to establish an average time scale for the process.

It has to be noted that $\Delta s$ can be either positive or negative. A negative value merely means that the particle is actually being accelerated or decelerated in the direction opposite to the field (acceleration will occur for a negative momentum wave vector). The value of $\Delta t$ should not depend upon this sign, but the change in momentum certainly does depend upon the sign. Thus, the momentum change given above should actually account for this sign as

$$\Delta k = \frac{eF\Delta t}{\hbar} sign(\Delta s) \ . \qquad (35)$$

One of the first steps to solving the Green's functions of sub-section 4.2 is determining the less-than self-energy, especially the imaginary part that is necessary in (30). This is done within the EMC procedure and this scattering function is shown in figure 5 for the same three values of electric field used earlier. In contrast to the retarded self-energy, in this case the less-than self-energy clearly is a function of the applied electric field, with more scattering at higher electric fields. This presumably is a result of more carriers at higher energies, where the scattering rates are higher, as shown in figure 1. This is also reflected in the shorter drift lengths apparent at higher values of the field in figure 4, although the higher field itself affects these lengths through the parameter $L$.

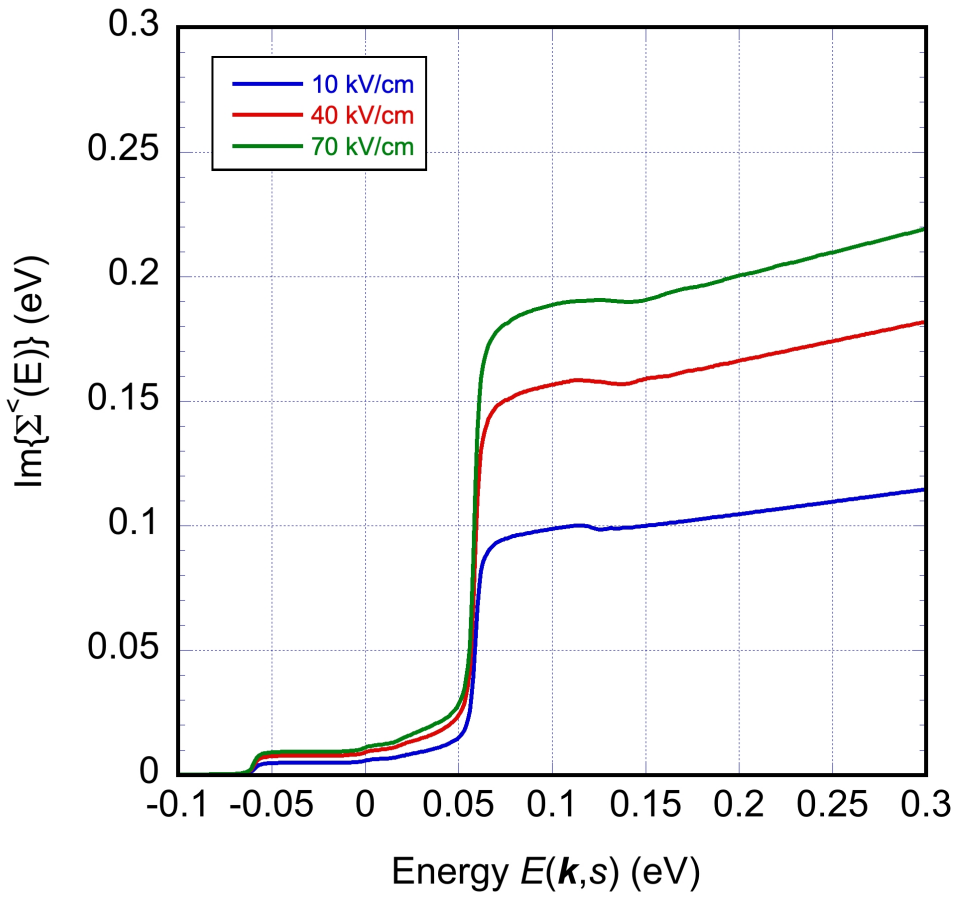

**Figure 5.** The imaginary part of the less-than self-energy is shown for three values of the electric field.

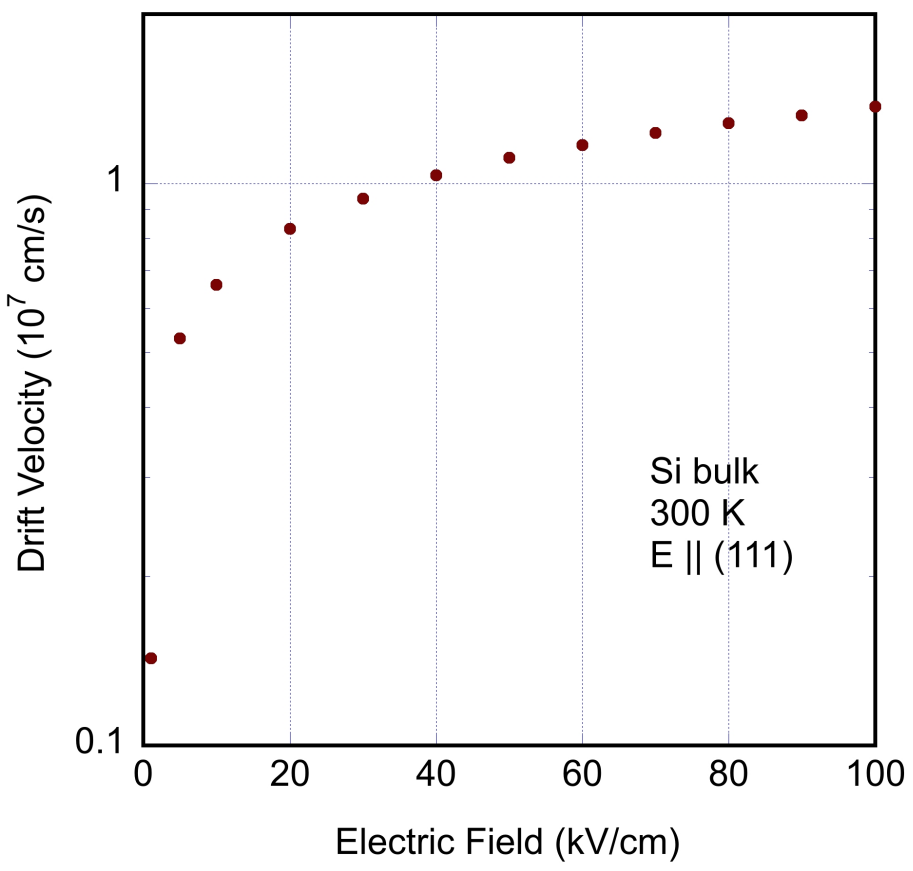

**Figure 6**. The drift velocity as a function of the electric field in Si at 300 K. This is determined from the ensemble averages and an average over the iterations.

The drift velocity is the average velocity of the ensemble of carriers, and is always computed by such an ensemble average. This ensemble average is then averaged over the 200 iterations of the simulation. The resulting average drift velocity is plotted as. a function of the electric field in figure 6. While they are not evident, error bars are actually shown in the figure, for the average over the iterations. These errors turn out to be less than 0.8% of the actual velocity, and so are buried under the symbol used in the plot. It might seem natural to determine the average velocity from the ration $\Delta s/\Delta t$. But, this would be a mistake as the latter quantity is the average over the drift length, and not the velocity at the end of the drift which is the quantity that leads to the drift velocity by virtue of Hamilton's equations of motion. The average of the velocity over the iterations is used to reduce the noise, but in many cases that can be misleading when the actual velocity is strongly time dependent. This is not the case here. The variation in the velocity during the simulation can be illustrated in another manner by plotting the velocity versus the average time that is determined from the average $\Delta t$ values, and this is done in figure 7. Each value of the field has its own apparent time scale, which is the average steps per iteration. As may be seen from figure 4 and (34), these time scales differ for each value of the electric field. In essence, it is perhaps better to consider these "time" scales as merely a parameter that measures the evolution of the ensemble over time. This differs from classical Monte Carlo, where the time is determined more or less accurately in the simulation. In this quantum version, it is the distance that is determined more or less accurately (we return to this point in the next section). However, after an initial value (after the first iteration) that is slightly higher than the subsequent values, the velocity is relatively constant. The noise may be estimated by the fluctuations in figure 7. The slightly higher value after the first iteration may be indicative of some velocity overshoot [64], but this cannot be determined properly, especially as the normal overshoot in Si is relatively small.

While the common value expected for velocity saturation is around $10^7$ cm/s, a higher value is found here (figure 6). It is felt that the lower value found in semi-classical Monte Carlo arises from the rapid energy rise in scattering by the first-order coupled low energy intervalley phonon. However, in the quantum case, it is found that this first-order scattering is especially affected by the intra-collisional field effect that arises in quantum transport [65]. Hence, it is not surprising that a larger "saturated" velocity is found in quantum transport.

In figure 8, the carrier population is plotted as a function of carrier energy for three different values of the electric field (same values as used earlier). As the field increases, carriers at low energy are lost to higher energy states, so that the distribution streams to higher energies. The various bumps and plateaus that tend to form arise from the tendency to have two-dimensional quantization and sub-bands in the high fields, as well as the role of onset of various phonons. The presence of particles at negative energies is the result of the broadening of the energy states, as represented by the spectral density of figure 3. While the plot uses relative units (the actual values do not relate to the number of particles), the differences between the three curves are accurate, and reflect the field induced differences.

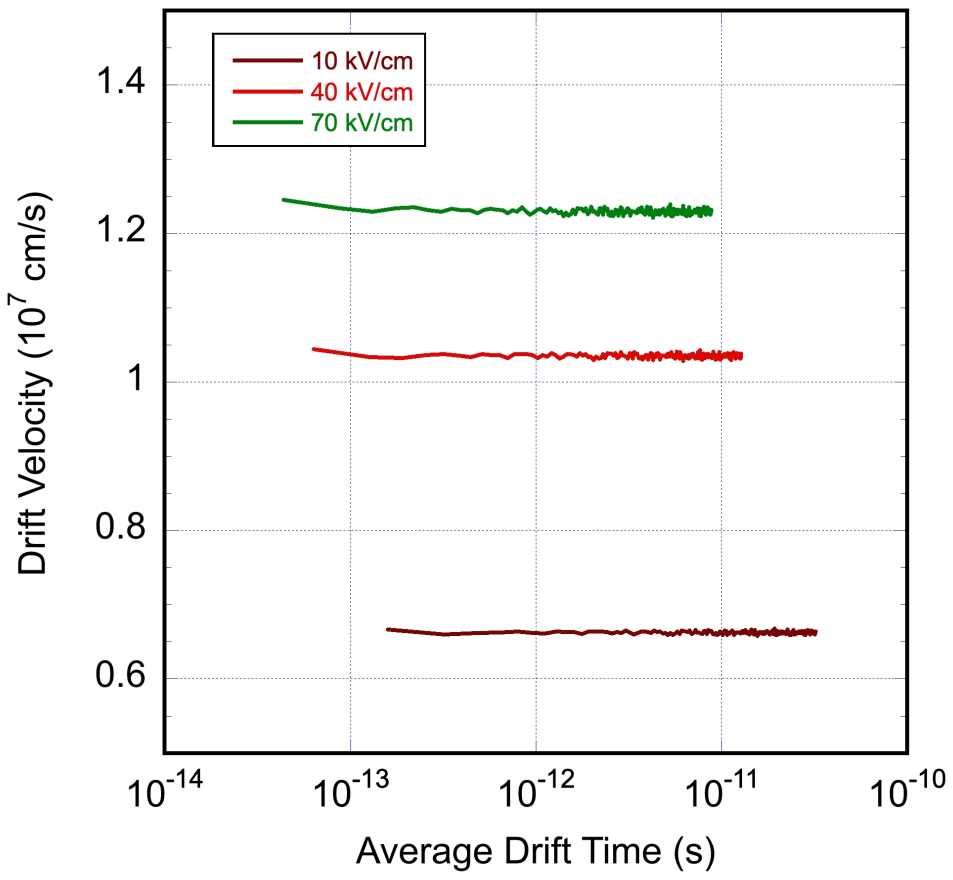

**Figure 7**. The variation in the drift velocity for three values of the electric field, as a function of the simulated time. This time scale differs for each value of the field, as discussed in the text.

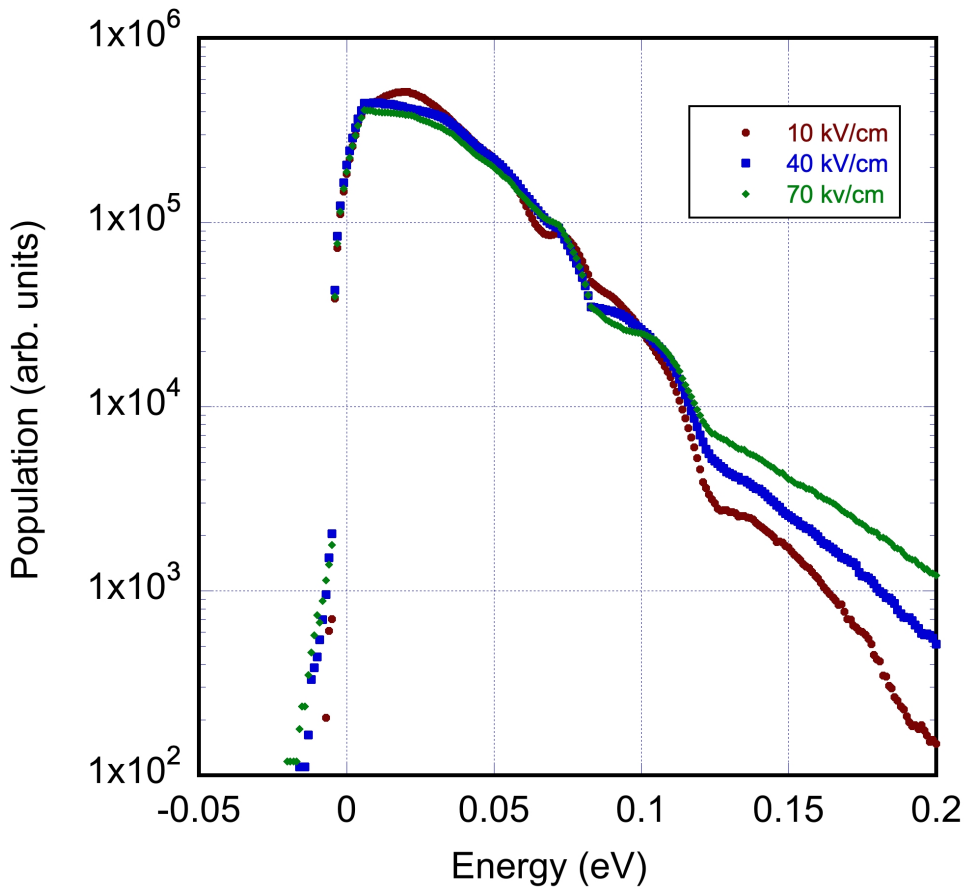

**Figure 8**. The relative carrier population as a function of energy. The bumps and plateaus are reflective of the tendency to two-dimensional quantization and some phonon effects.

In figure 9, the distribution functions determined here are compared to a thermal Maxwellian in panel (a), and with the Airy Green's function approach of Bertoncini and Jauho [66]. In panel (a), it is apparent that the form of the distribution has begun to deviate from the thermal one, with the need for the distribution to approach a balance between the driving force (the electric field) and the dissipative scattering processes. this usually leads to a non-thermal distribution for hot carriers [67]. In panel (b), the distribution is compared to the previous work for a field of 100 kV/cm. The present result does not stream to as high energies as the earlier work of Bertoncini and Jauho. This is due to the latter using only a single high energy phonon in the simulation. It is know from classical simulations that the first-order coupled phonons are necessary to bring the velocity down to the observed $\sim 1.1 \times 10^7$ cm/s [61,67,68] and it may be seen in their work that a classical simulation gives a velocity of $2 \times 10^7$ cm/s, well above the experimental results [68]. So, it is likely that the additional inclusion here of the first-order optical modes and the acoustic modes lead to less streaming that seen in the earlier work.

## 6. Discussion

In this paper, the use of the ensemble Monte Carlo process has been shown to be effective in evaluating non-equilibrium Green's functions in a fast and effective approach. While this procedure is limited, at the present time, to materials in which the scattering is very local, such as with nonpolar optical and acoustical phonons, this application is of importance in most semiconductor devices used for integrated circuits. Typically, the computation for a single value of the electric field takes only on the order of 1.6-2.1 seconds on a modern laptop computer (here a MacBook Pro with the M1 Pro chip), with parallel processing available.

The results obtained for the drfit velocity in figure 6 are intriguing, in the sense that above ~40 kV/cm, there does not appear to be any hard saturation of the velocity, which is commonly seen in semi-classical Monte Carlo approaches [69,70], with a value of $1 \times 10^7$ cm/s at room temperature, although some versions give values well above this [71]. However, this can be misleading, as there are a wide variety of such approaches. However, measurements are not clear on this point. Most measurements do not go beyond 40-50 kV/cm. For example, the oldest measurements using time-of-flight techniques extend to approximately 50 kV/cm [68], 13 kV/cm [72], or 20 kV/cm [73], so that the appearance of a hard saturation is not seen in the data. It has also been found that quantum simulations generally give higher values of velocity than the semi-classical ones [66,74]. The value of $1.3 \times 10^7$ cm/s found here is close to that found by Reggiani *et al*. [71].

If one accepts that the quantum drift velocity is larger than that found semi-classically, there must be a reason. First, it has been known for some time that the collision broadening inherent in the spectral density contributes to this [71]. In addition, normally

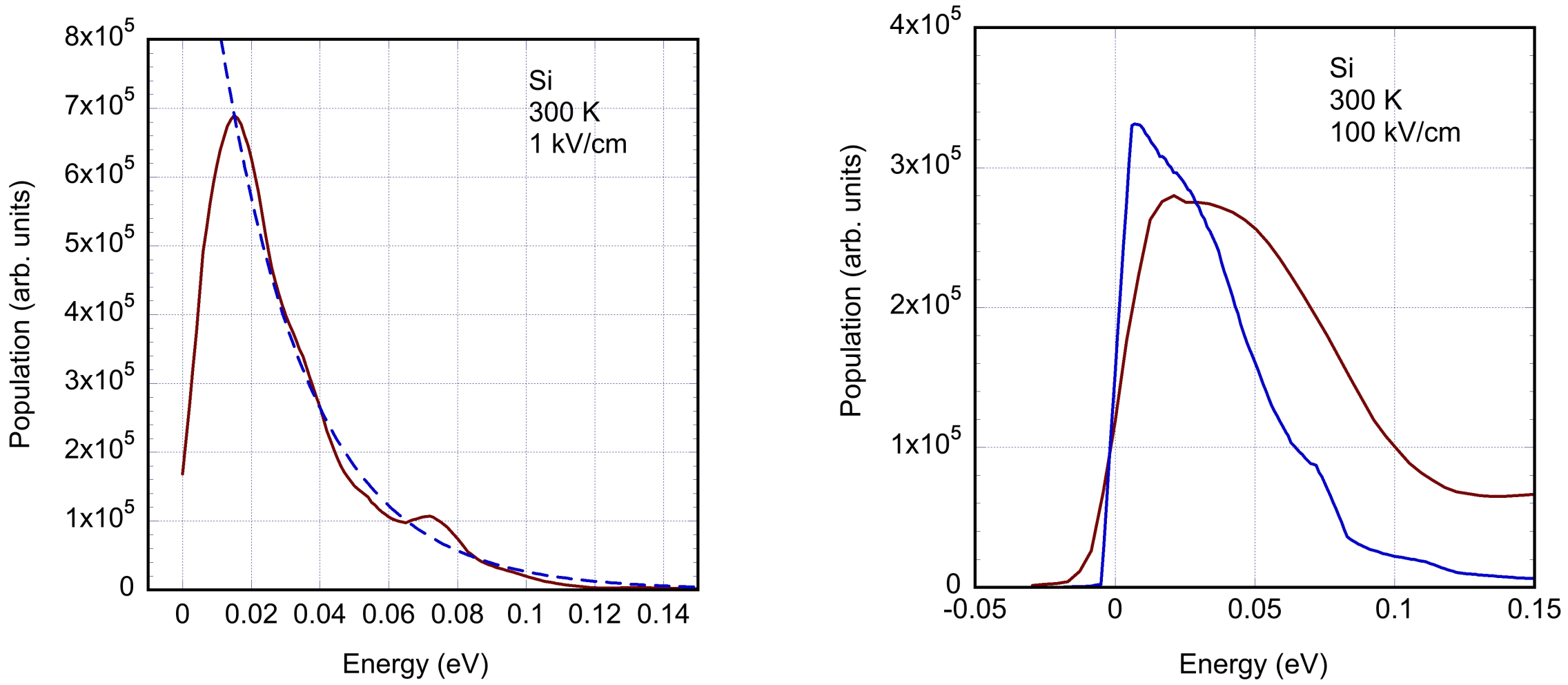

(a) (b)

Figure 9 (a) Comparison of the carrier distribution function with a classical, thermal distribution at 1 kV/cm. (b) Comparison of the distribution function computed here with that of ref. [66]. These are discussed in the text.

the Monte Carlo determines the ballistic drift length from a random time step derived from the total scattering rate. In the quantum method presented here, the ballistic drift length is determined solely by the distance step that depends only upon the electric field. Moreover, the important scattering arises from the imaginary part of the less-than self-energy (figure 5) and is this directly a function of the electric field. Thus, there is a direct inter-twining of the field and the scattering that contributes to the effect referred to as the intra-collisional field effect [3,75], a point emphasized earlier [71]. But, there is a further contribution, apparent in (30). This is the fact that the final states for scattering are an almost two-dimensional density of states, represented by the inverse hyperbolic tangent function. Quite generally, the density of states in a two-dimensional system is less than that in a three-dimensional system. This lower density of states gives less scattering and therefore a higher velocity. So there are several reasons for the drift velocity to be higher in quantum simulations than in classical ones.

Another important consideration for the future is that the present approach has been applied for homogeneous material, while devices tend to be quite inhomogeneous. Extending this approach to inhomogeneous devices should be only a minor problem, as the Monte Carlo technique is quite usual in such device simulation. Most of the computational time is taken with Poisson's equation, and from the results of this equation, the spatially dependent quantum distribution can be developed rather quickly.

In devices, there is usually a grid upon which the density is projected in order to solve Poisson's equation. Moreover, when Monte Carlo techniques are used for the transport in cases of real-space treatment of the impurities [76], or treating the electron-electron interaction by molecular dynamics [77], there are usually two time scales, one attached to each particle and the laboratory time scale for updating the total potential. Techniques have been adopted for synchronizing these two times and distributing the acceleration between multiple bins of the laboratory time scale, since the potentials have to be updated on a much faster time scale than the scattering events [58]. In principle, dividing the $\Delta s$ or $\Delta t$ values can easily be done by this same approach for either inhomogeneous material and devices or for incorporation of carrier-carrier interactions.

Extending this approach to the polar optical phonons involves the problem of the long-range of Coulomb interactions. The polar interaction is a dipolar one, but still is quite inhomogeneous in the scattering dynamics just as is impurity scattering. The latter has been made more amenable through treating the impurities in real space and developing the total potential from this real space approach [76]. This has not been done for the polar modes as yet. For solutions with the Green's functions, this should require direct solution of the Bethe-Salpeter equation for the very inhomogeneous scattering by the polar phonons. However, this latter equation is itself an iterative procedure for high accuracy, and it is not beyond expectations that a second Monte Carlo process may be used for its evaluation, beyond the Monte Carlo procedure for determining the resultant Green's function and distribution function. It is also feasible to consider that the polar mode interaction may be

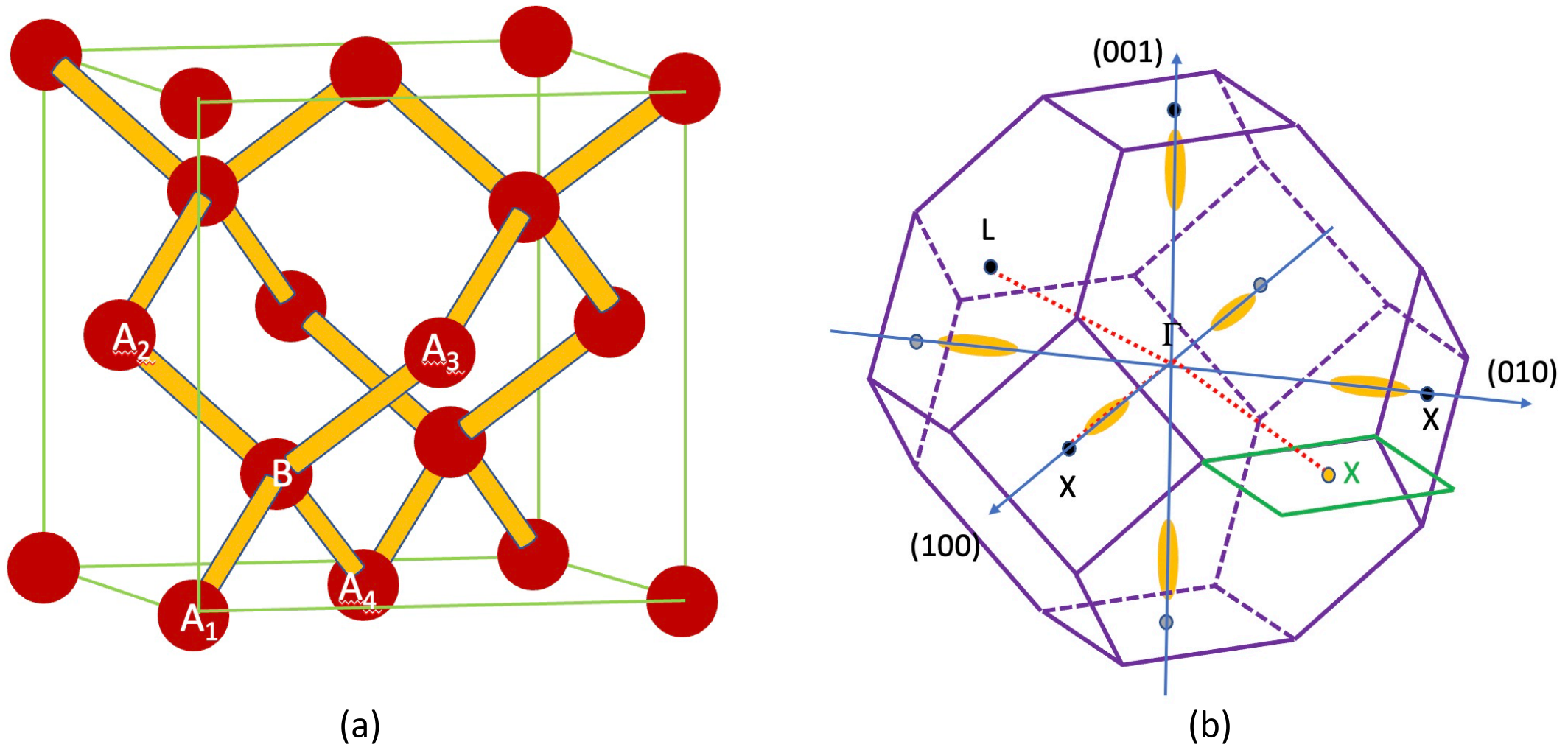

**Figure A1** (a) The diamond structure, in which Si crystallizes, is a face-centered cube, as shown here. (b) The first Brillouin zone is depicted here. The solid lines are foreground, while the dashed are background. The major axes are depicted with their index notation. The green square and gold ellipses are constant energy surfaces for an energy near the conduction band minima. The dashed red lines are discussed in the text.

expressed in a real space approach, such as used for the impurities. While these suggestions remain to be speculation, it is certainly worth further research in this area to extend the present approach to devices and polar materials.

**Appendix: Basic Properties of Silicon**

As mentioned above, Si is perhaps the most used semiconductor in the electronics world and is the base material in nearly all integrated circuits (or "chips"). Si crystallizes in the diamond structure, which is a face-centered cube with a basis of two atoms, as shown in figure A1(a). The basis includes one atom, noted as $A_1$, and its partner denoted as B in the figure. The atoms are tetrahedrally coordinated, and the four neighbors of the B atom are the four denoted with $A_i$ $(i = 1 - 4)$. These four A atoms form a regular tetrahedron. The bonds are all covalent in nature. The face-centered cube is not the primitive (or smallest periodic structure) unit cell. The latter contains only one A atom and one B atom, and is defined by vectors from $A_1$ to the other three A atoms. There are four such primitive cells in the face-centered cube. The edge length of the face-centered cube is denoted as $a$, and has a value of 0.543 nm. The distance between the A and B atoms is $\sqrt{3}a/4 = 0.235$ nm.

In figure A1(b), the first Brillouin zone of the reciprocal lattice (momentum space) is depicted. This is a cube in which the eight corners have been truncated (cut off). The principle axes are shown for $k_x, k_y, k_z$ and are denoted by their Miller indices as (100), (010), and (001), respectively [78]. These points lie $\pi/a$ from the origin, which is denoted as Γ. The gold ellipses are constant energy surfaces for an energy near the conduction band minima. There are six such ellipses due to the six-fold cubic symmetry of the set of (100) axes.

When the conduction band is computed, it is quite complicated to plot the entire 3D energies in a viewable manner. As a consequence, one usually plots along a set of symmetry lines [79]. Here, the various energy bands will be plotted along a line from L (the center of one of the hexagonal faces) along a (111) line to the center point Γ and then along (100) to the zone edge at X. A shift to the second zone X (in green) is made and the plot proceeds from here back to Γ along the (110) line shown in dotted red. When this line crosses into the first zone, the point is often denoted as W or U. When the various Brillouin zones are stacked, they are connected at the hexagonal faces. Thus, a second zone has its square face at (001) adjacent to the edge depicted by the green square. Thus, the point X shown in green in this square is in the second Brillouin zone. The resulting band structure, as computed by a nonlocal empirical pseudopotential method [80] is shown in figure A2. Here, only a range of energies around the principal band gap is shown. The top of the valence band lies at Γ while the minima of the conduction band lie about $0.85\,\pi/a$, or about 85% of the distance to X. Constant energy surfaces, for an energy slightly above these minima are shown as green ellipses in figure A1(b).

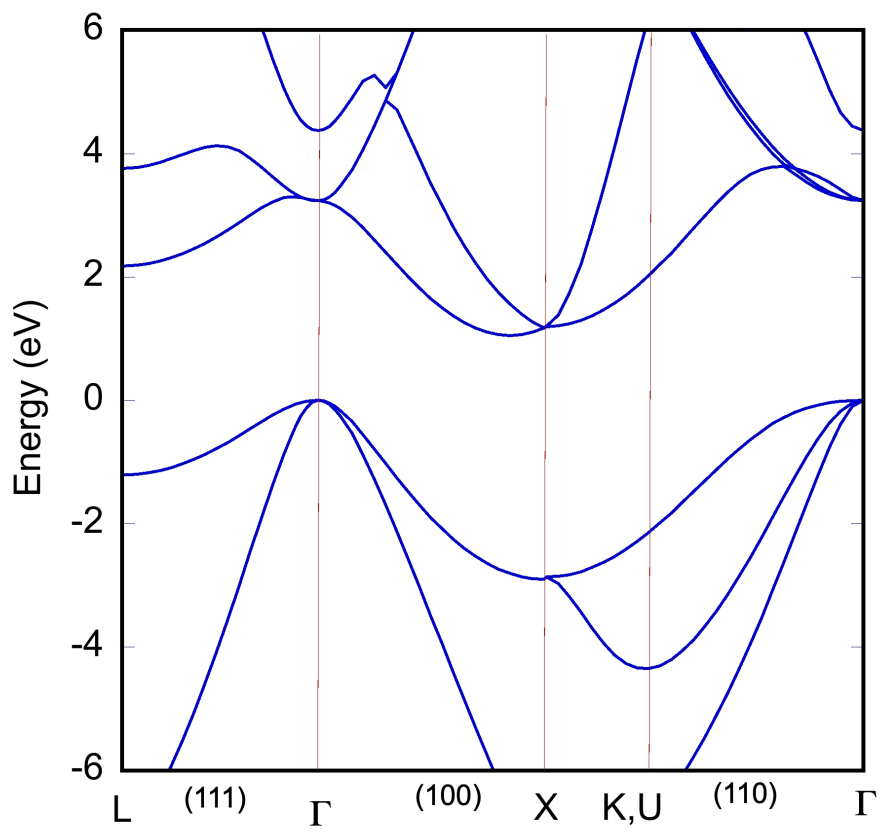

**Figure A2** The conduction and valence bands for Si are plotted along a set of axes in the Brillouin zone. The path followed is discussed further in the text.

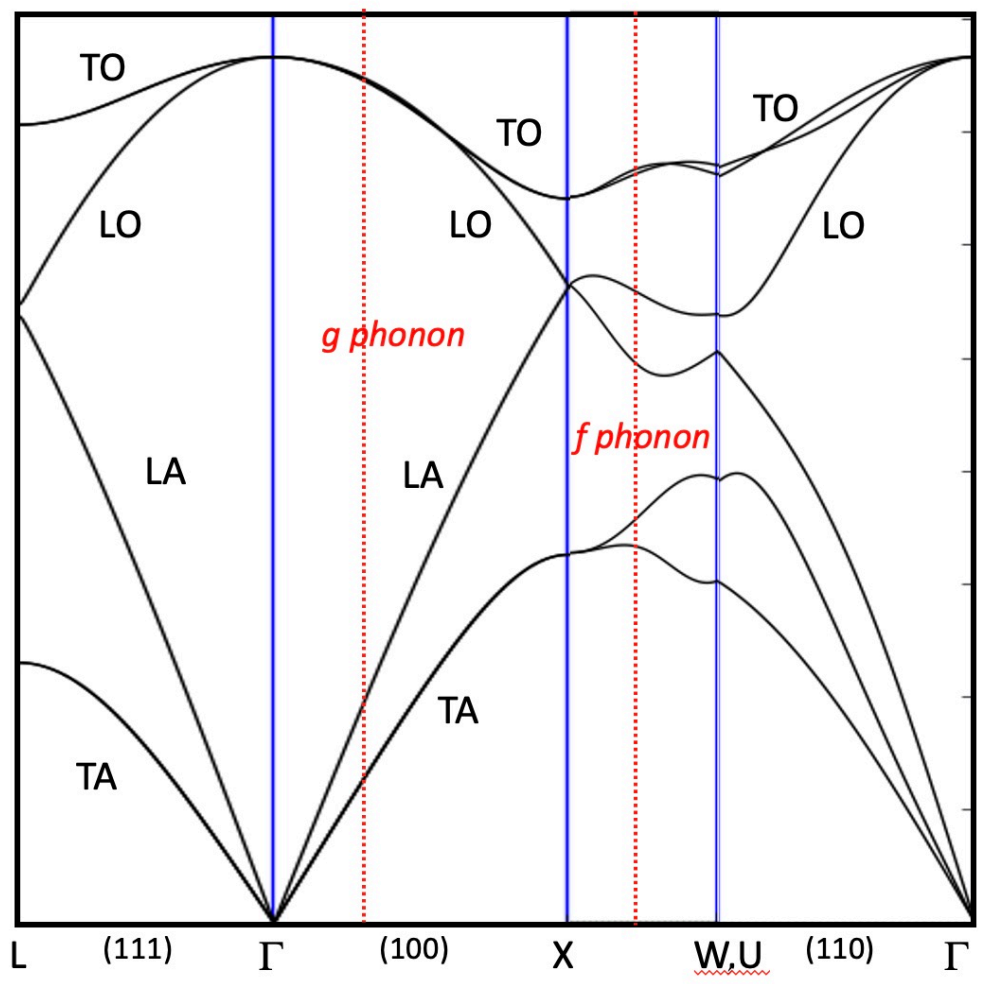

**Figure A3** The phonon bands are shown for the Si crystal along the same path in the Brillouin zone as used for the band structure of figure A2. The various phonons used for electron scattering between the various ellipses are denoted by the two vertical red lines, discussed further in the text.

The vibrations of the atoms are generally discussed in terms of their various modes in momentum space. As there are two atoms in the basis, and in the primitive unit cell, there are six various modes in three dimensions. Only three of these can be acoustic modes, in which both atoms in the primitive cell move together as a pair. The other three modes create the optical modes, in which the two atoms move in opposite directions [58]. These phonon bands are shown in figure A3, along the same path in the Brillouin zone as the energy bands shown in figure A2. The various bands are labeled by their nature--L and T for longitudinal and transverse (for whether the motion is parallel or perpendicular, respectively, to the propagation direction) and A and O for acoustic and optical. The very low energy acoustic modes (near Γ) mainly scatter within one of the ellipses of the conduction band, while the important scattering between the ellipses is done by higher energy modes.

As discussed briefly in sub-section 4.3, there are two sets of phonons that lead to scattering between the ellipses of the conduction band. One set is labeled as $g$ phonons and has a momentum vector that stretches between the two ellipses along a common direction, such as between the (100) and $(\bar{1}00)$ (where the bar over the top denotes a negative value). This vector thus has a length of $2 \times 0.85\,\pi/a$ and extends outside the first Brillouin zone. The periodicity in momentum space means that any vector such as this is equivalent when a reciprocal lattice vector is added or subtracted. Thus, if we reduce the momentum by the reciprocal vector $2\,\pi/a$, then the equivalent vector is $0.3\,\pi/a$, and is shown by the vertical red line in figure A3 (identified as the $g$ phonon in the figure). The scattering between ellipses on different lines, such as between (100) and (010) are denoted as $f$ phonons, and have a momentum vector whose length is $\sqrt{2} \times 0.85\,\pi/a \sim 1.2\,\pi/a$. Once again, this vector lies outside the first Brillouin zone, and falls in the square face of the second zone, shown in green in figure A1(b). This position is shown by the vertical red line, designated as $f$ phonon in figure A3. In each case, there are optical and acoustic modes that can span these distances. The acoustic modes are generally forbidden in a zero-order interaction (no $q$ vectors in the matrix elements; even the low energy acoustic modes are actually first-order, having a $q$ vector in the matrix element). The optical mode interaction is of zero order, but experimental results suggest that the acoustic modes do occur in these interactions [59], so that they obviously require a first-order calculation [61,62]. And, this is achieved in the current simulations as well as in semi-classical interactions [67].